\documentclass{article}
\usepackage{amsfonts, amssymb,amsmath} % Include mathematical symbols
\usepackage{graphicx}% Include figure files
\usepackage{dcolumn}% Align table columns on decimal point
\usepackage{bm}% bold math
\newcommand{\be}{\begin{equation}}
\newcommand{\ee}{\end{equation}}
\newcommand{\bea}{\begin{eqnarray}}
\newcommand{\eea}{\end{eqnarray}}

\newcommand{\ti}{\i}

\begin{document}

%\preprint{APS/123-QED}
\begin{center}
{\Large A New Method for Derivation of Statistical Weight of the
Gentile Statistics}
\\
\vspace{.3cm}
Sevilay Selvi, Haydar Uncu \\
Department of Physics, Adnan Menderes University, Aytepe, 09100,
Ayd\ti n, Turkey \end{center}
%\date{\today}
%-------------------------------------------------------------------------------
\begin{abstract}
We present a new method for obtaining the statistical weight of
the Gentile Statistics. In a recent paper, Perez and Tun presented
an approximate combinatoric and an exact recursive formula for the
statistical weight of Gentile Statistics, beginning from bosonic
and fermionic cases, respectively \cite{pereztun}. In this paper,
we obtain two exact, one combinatoric and one recursive, formulae
for the statistical weight of Gentile Statistics, by an another
approach. The combinatoric formula is valid only for special
cases, whereas recursive formula is valid for all possible cases.
Moreover, for a given q-maximum number of particles that can
occupy a level for Gentile statistics-the recursive formula we
have derived gives the result much faster than the recursive
formula presented in \cite{pereztun}, when one uses a computer
program. Moreover we obtained the statistical weight for the
distribution proposed by Dai and Xie in Ref. \cite{xie}.
\end{abstract}

---------------------------------------------------------------------
%------------------------------------------------------------------------
%\maketitle
%\narrowtext
%==============================================================================
% Section 1
\section{Introduction \label{intro} }
 An interesting property of low dimensional systems is that the
particles in these systems may obey different statistics other than
Bose-Einstein and Fermi-Dirac statistics \cite{avinashkhare}
\footnote{There is a huge literature about the intermediate
statistics one can cite. For this reason, we have chosen to cite a
book.}. Following the realization that there can be quasi particles,
whose many body wave function may have a general phase $e^{i
\theta}$ \cite{leinaas,wilczek,nayak} -other than $1$ or $(-1)$-,
Haldene proposed a fractional statistics in arbitrary dimensions
\cite{haldene}. Then, Wu derived statistical weight for the Haldene
fractional statistics \cite{wu}
\begin{equation}
W_{i}=\frac{[g_{i}+(n_{i}-1)(1-\alpha)]!}{n_{i}![g_{i}-\alpha
n_{i}-(1-\alpha)]!} .\label{whaldene}
\end{equation}
Here, $n_i$ gives the identical number of particles occupying a
state $i$ and $g_{i}$ is the number of states. The parameter
$\alpha$ in Eq.\eqref{whaldene} yields an interpolation between
Bose-Einstein and Fermi-Dirac statistics. So, the statistical weight
Wi in $(1)$ reduces to Bose-Einstein and Fermi-Dirac statistics, for
$\alpha=0$ and $\alpha=1$, respectively. On the other hand,
Polychranakos suggested another form for the statistical weight of
the Haldene fractional statistics \cite{polychranokos}.

The possibility of intermediate statistics led to the new studies on
Gentile Statistics, which was proposed much before than the other
generalizations of the Bose-Einstein and Fermi-Dirac statistics
\cite{gentile}. For example, Dut et. al have shown that the
expressions for distribution and other thermodynamic quantities
derived using Gentile statistics are also valid for a q-fermion
provided q is a complex number and takes values on a unit circle
\cite{dutt}. Then, Chaturvedi and Srinivasan compared different
interpolations between Fermi and Bose Statistics including Gentile
statistics \cite{chaturvedi}. Bysto derived a thermodynamic Bethe
ansatz equation for relativistic particles obeying generalized
extensive statistics \cite{bysto}. Moreover, Dai and Xie showed that
Gentile statistics does not reduce to Bose-Einstein statistics
generally but only if fugacity $z=e^{\beta\mu}<1$ \cite{daiannals}.
In addition to that, they show that one can obtain Bose-Einstein
distribution $f_{BE}$, from Gentile distribution $f_{G}$, in
thermodynamic limit when one takes two limits, maximum occupation
number $q\rightarrow\infty$, and the total number of particles
$(N\rightarrow\infty)$ in a given order $(f_{BE}\equiv  \lim_{<N>
\rightarrow \infty}{\lim_{n \rightarrow \infty}f_{G}})$
\cite{daiphysics}. Furthermore, Donald and Zly derived thermodynamic
properties for a harmonically confined gas obeying Gentile
statistics in d-dimensions and compared these results with a similar
system obeying Haldene-Wu statistics \cite{macdonald}.

There are also rather mathematical studies about Gentile statistics.
For example, the relationship between Gentile statistics and
restricted partitions is investigated by Srivatson et. al.
\cite{sriatson}. Moreover, Niven studied the combinatorial entropies
and statistics \cite{niven}, and Mirza and Mohommadzeh investigated
thermodynamics geometry of fractional statistics \cite{mirza}.

It is also possible to obtain intermediate statistics from
operator relations. For example Melijenac et. al. studied
exclusion statistics in the second quantized approach which
includes Gentile statistics as a special case \cite{melijanac}. In
addition to that, Dai and Xie obtained an operator realization for
the angular momentum algebra which naturally leads to Gentile
distribution \cite{xie2}. Whet significant in this derivation is
that one does not need to restrict the number of the particles by
fiat as it is done in the Holstein-Primakoff representation
\cite{HP} because it arises naturally in this representation
\cite{xie2}.  Then they applied this distribution to the
excitations of the spin magnetic waves for the one dimensional
Heisenberg chain and showed that the distribution they have
obtained explains the excitation spectrum better than the
Hollstein-Primakoff method \cite{xie2}. Moreover, the same authors
derived Gentile statistics from operator relations \cite{xie1}. In
this study, the authors used an algebra similar to that of the one
dimensional harmonic oscillator algebra but in this algebra
creation and annihilation operators are not hermitian conjugate of
each other. Using this more generalized algebra, they showed that
in the algebras where a number operator $\hat{N}$ can be defined a
quadratic function of creation and annihilation operators one gets
the Gentile distribution corresponding to this algebra
\cite{xie1}.

Recently, several studies showed that, Gentile statistics can also
be used to describe realistic physical systems. Indeed, Gentile
statistics is appropriate for investigating two dimensional
electron gases in two dimension when the electrons are so dilute
that the Coulomb interactions between them is negligible. In this
case, the behavior of the electrons is described by a Hamiltonian
similar to the harmonic oscillator hamiltonian but having one more
degree of freedom. Therefore, more than two electrons (including
spin degeneracy) may occupy each single particle states but the
maximum  number of the electrons in a state is limited by the
magnetic field leading to the de Haas-van Alphen effect
\cite{grosso}. The statistical distribution of these electrons is
the Gentile distribution. Moreover, Auccaise et.al. presented a
description of nuclear magnetic resonance of quadrupolar system
\cite{auccaise}. In this work, using Holstein-Primakoff
representation \cite{HP} the authors proposed that NMR quadrupolar
system have BEC like behavior and they experimentally verified
their results for two different quadrapol nuclei ($^{23}Na$) and
($^{133}Cs$) in lyotropic cyrstals which have nuclear spins
$I=3/2$ and $I=7/2$, respectively. These system are interesting
because their statical properties may be obtained using the
Gentile statistics\cite{auccaise}. Moreover, Shen and Yin showed
that cyclich hydrocarbon polyenes $C_NH_N$, called N-annules, are
physical realizations of Gentile systems \cite{yoshen}.

In this paper, we present a new inductive method to obtain the
statistical weight of the Gentile statistics. The statistical
weight for Gentile statistics is first derived by Perez and Tun
\cite{pereztun}. We will derive two formulae one combinatoric and
one recursive like in \cite{pereztun}. Both formulae are exact but
the combinatoric one is valid only for special cases. Since, the
statistical weight for Gentile statistics is derived earlier in
\cite{pereztun}, we owe to an explanation why we have done this
study. We obtain an exact combinatoric formula  for all q values,
maximum number of particles which can occupy a state, which
however is only valid for $(G-1)q \leq N \leq Gq$ , where $N$ is
total number of the particle and $G$ is the number of different
states, respectively. The combinatoric formula obtained in
\cite{pereztun} is valid only for $q \geq N/2$. The recursive
formulae are useful only when one uses computers, and the
recursive formula we obtained produces the statistical weight much
faster, compared to the recursive formula obtained in
\cite{pereztun},  when $q$ has a determined value like in $N$
annules \cite{yoshen}. Moreover, the inductive method constructed
in this work can be applied to find the statistic weight
constructed in \cite{xie} by Xie and Dai which is more general
than Gentile statistics. We will denote this statistics as Dai-Xie
distribution.
\section{Method}
In this section, we will present a new inductive method for
obtaining the statistical weight of Gentile statistics. This method
can also be used to obtain the statistical weight for Bose-Einstein
statistics. For the sake of simplicity, we first apply the method
for obtaining the statistical weight of Bose-Einstein statistics.

The statistical weight for bosons give the number of different ways
of distributing N bosons to G levels, $R(N,G)$. Since our method is
inductive, we first find how many different ways to distribute N
bosons to two levels. (Obviously, there is only one way to
distribute N bosons to one level.)
\begin{figure}[b!]
\hskip 1.9cm
\includegraphics[height=4cm, width=8cm]{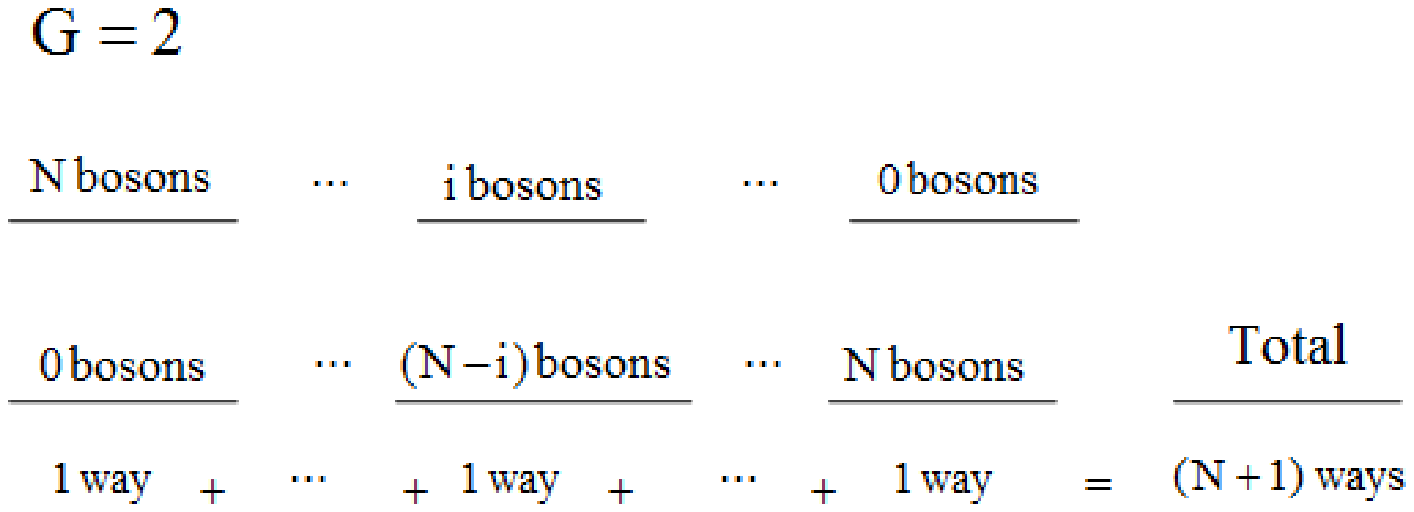}
\caption{Statistical weight of N bosons for two levels $R(N,2)$}
\label{bozonikia}
\end{figure}

\begin{figure}[t!]
\hskip 1.9cm
\includegraphics[height=6cm, width=8cm]{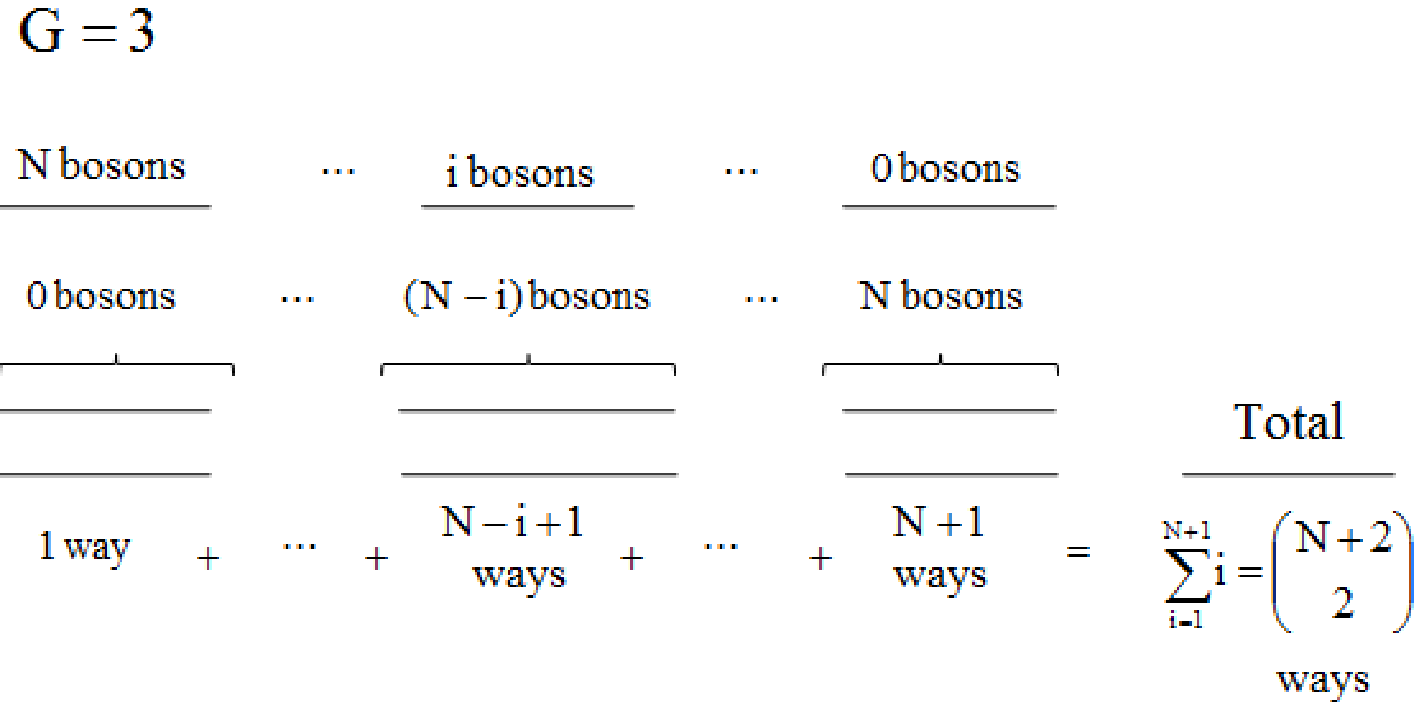}
\caption{Statistical weight of N bosons for three levels $R(N,3)$}
\label{bozonuca}
\end{figure}

In order to find the number of ways to distribute N bosons to two
levels $R(2,N)$, we count all different cases (see Figure
\ref{bozonikia}). There can be $0$ bosons in level $2$ and all
bosons can be level $1$. Or, there can be $1$ boson in level $2$ and
$N-1$ bosons in level $1$, $2$ bosons in level $2$ and $N-2$ bosons
in level $1$ and so on. Counting the different cases, we find that
there are $(N+1)$ ways to distribute $N$ bosons two levels. Now, we
will find $R(3,N)$ using $R(2,N)$. We again use the same logic. We
assume first there are $0$ bosons in level 1. Thus there are $N$
bosons in level $2$ and $3$. Since we know $R(2,N)=N+1$, we conclude
if there are $0$ bosons in level 2, then three are $(N+1)$ ways to
distribute $N$ bosons to there levels (see Figure \ref{bozonuca}).
If, there is only one boson in level two, there are $N-1$ bosons in
level $2$ and $3$. Therefore, in this case, there are $R(2,N-1)=N$
different ways to distribute $N$ bosons to three levels. By using
the same logic one can conclude, if there are two bosons in level
$1$ there are $N-1$, if there are three bosons in level $1$ there
are $N-2$ ways to distribute $N$ bosons to three levels and so on.
Generally, if there are $i$ bosons in level $1$, there are $(N+1-i)$
ways to distribute $N$ bosons to three levels. Since there can be at
least $0$ bosons and at most $N$ bosons in level $1$, the total
number of ways to distribute $N$ bosons to three levels is
\begin{equation}
R(3,N)=\sum_{i=1}^{N}(N+1-i)=\sum_{j=1}^{N+1}j=\frac{(N+1)(N+2)}{2}=
\left( \begin{array}{c} N+2
\\ 2
\end{array} \right).
\label{bozonuc}
\end{equation}
Similarly, it is easy to find $R(4,N)= \left( \begin{array}{c} N+3
\\ 3
\end{array} \right)$. Using results for $R(3,N)$ and $R(4,N)$ we
infer a general formula to distribute N bosons to G Levels:
\begin{equation}
R(G,N)=\left( \begin{array}{c} N+G-1
\\ G-1
\end{array} \right).
\label{bozonN}
\end{equation}
Now, we have to show that $R(G+1,N)=\left(
\begin{array}{c} N+G
\\ G
\end{array} \right)$ assuming Eq. \eqref{bozonN}
is valid to complete the induction. To do this, we use the method we
have used to find  $R(2,N)$ and $R(3,N)$. We first assume that there
$0$ bosons in level $1$. Thus, there are N bosons in the remaining G
levels and $R(G,N)=\left(
\begin{array}{c} N+G-1
\\ G-1
\end{array} \right)$ different ways to
distribute $N$ bosons to these levels. If there are $1$ boson in the
first level, then there are $R(G,N-1)=\left(
\begin{array}{c} N-1+G-1
\\ G-1
\end{array} \right)$ ways to distribute
remaining $N-1$ bosons to the G groups. Continuing this process and
summing the number of all different ways for the different numbers
of bosons in level $1$ we get
\begin{equation}
R(G+1,N)=\sum_{i=0}^{N}\left( \begin{array}{c} N+G-1-i
\\ G-1
\end{array} \right),
\nonumber
\end{equation}
\begin{equation}
\sum_{k=G-1}^{N+G-1}\left( \begin{array}{c} k
\\ G-1
\end{array} \right)=\left( \begin{array}{c} N+G
\\ G
\end{array} \right).
\label{bozonNG}
\end{equation}
In order to get third term from the second one in Eq.
\eqref{bozonNG} we have changed the dummy index $i$ to $k=N+G-1-i$.
Moreover we have used the equality $\sum_{j=m}^n\left(
\begin{array}{c} j
\\ m
\end{array} \right)=\left( \begin{array}{c} n+1
\\ m+1
\end{array} \right)$ for $n \geq m$.
Since Eq. \eqref{bozonNG} is the same as the Eq. \eqref{bozonN} for
$G$ replaced by $(G+1)$, we conclude that our assumption namely Eq.
\eqref{bozonN} is valid.

The inductive method we have introduced in the previous paragraphs
can also be used to find the statistical weight for systems
obeying Gentile distribution (Gentile particles). For Gentile
particles there is an upper limit for the number of particles that
can occupy a level and we denote this limit by q. By using the
inductive method for Gentile particles one has to take this limit
into account.

We will now find the the statistical weight for Gentile particles.
Following the notation introduced in \cite{pereztun}, we will
denote the statistical weight for Gentile particles by $R_q(G,N)$
which shows the number of different ways to distribute $N$
particles into $G$ levels for a given q. The total number of
Gentile particles for given $G$ and $q$ has  an upper limit
$N_{max}=Gq$.

We first investigate the case $N \leq q$. In this case, $R_q(G,N)$
is the same as the boson distribution for all $G$. Since the limit
$q$ is greater than the total number of particles it does not effect
the occupancy of a level \footnote{Thus, one can think there may be
a very large limit for bosons to occupy a state.}. Therefore we can
write
\begin{equation}
R_q(G,N)=\left( \begin{array}{c} N+G-1
\\ G-1
\end{array} \right)\;\;\;\;\;\;\;\;\;\;N\leq q .
\label{Gentilebozonic}
\end{equation}
In order to find the statistical weight for all cases, we again
start with two levels,  i.e., we will first calculate $R_q(2,N)$. We
will separate the cases $q < N \leq 2q$ and $N\leq q$. Since for $q
\geq N$ the statistical weight for Gentile particles is the same as
the statistical weight for bosons, we get
\begin{equation}
R_q(2,N)=R(2,N)=(N+1) \qquad N \geq q \label{bozoniki}
\end{equation}
If $q < N \leq 2q$, we count the number of different ways,
separating level $1$. One can put minimum $N-q$ (otherwise there
would be more than $q$ particles in level $2$, which is not allowed
for Gentile particles), maximum q particles to level 1. Since there
is only one level left, there is only one way to distribute the
remaining particles to the remaining level. Therefore, there are
$2q+1-N$ number of ways to distribute $N$ particles to two levels if
$q < N \leq 2q$. Hence, we can write
\begin{equation}
R_q(2,N)= \begin{cases} 2q+1-N \;\;\;\;\;\; q \leq N \leq 2q
\\ N+1 \;\;\;\;\;\; N \leq q
\end{cases}
\label{goniki}
\end{equation}
The two equalities in Eq. \eqref{goniki} give the same number for
$N=q$. So one can use any of them for this case.

We will now try to find $R_q(3,N)$. Since the case $N \leq q$, is
boson distribution we will try to find $R_q(3,N)$ for $N>q$. We
separate cases $2q \leq N \leq 3q$ and $N < 2q$. Because the formula
for $R_q(2,N)$ differs for cases $q \leq N$ and $N<q$. If there are
more than $2q$ Gentile particles to distribute to three levels, the
total number of particles in the levels 1 and 2 has to be more than
$q$, because we can put at most $q$ particles to level $1$.
Therefore, one can see from Eq. \eqref{goniki} that the expression
for the statistical weight differ for $2q\leq N\leq 3q$ and for
$N<2q$.
\begin{figure}[t!]
\hskip 1cm
\includegraphics[height=6cm, width=8.8cm]{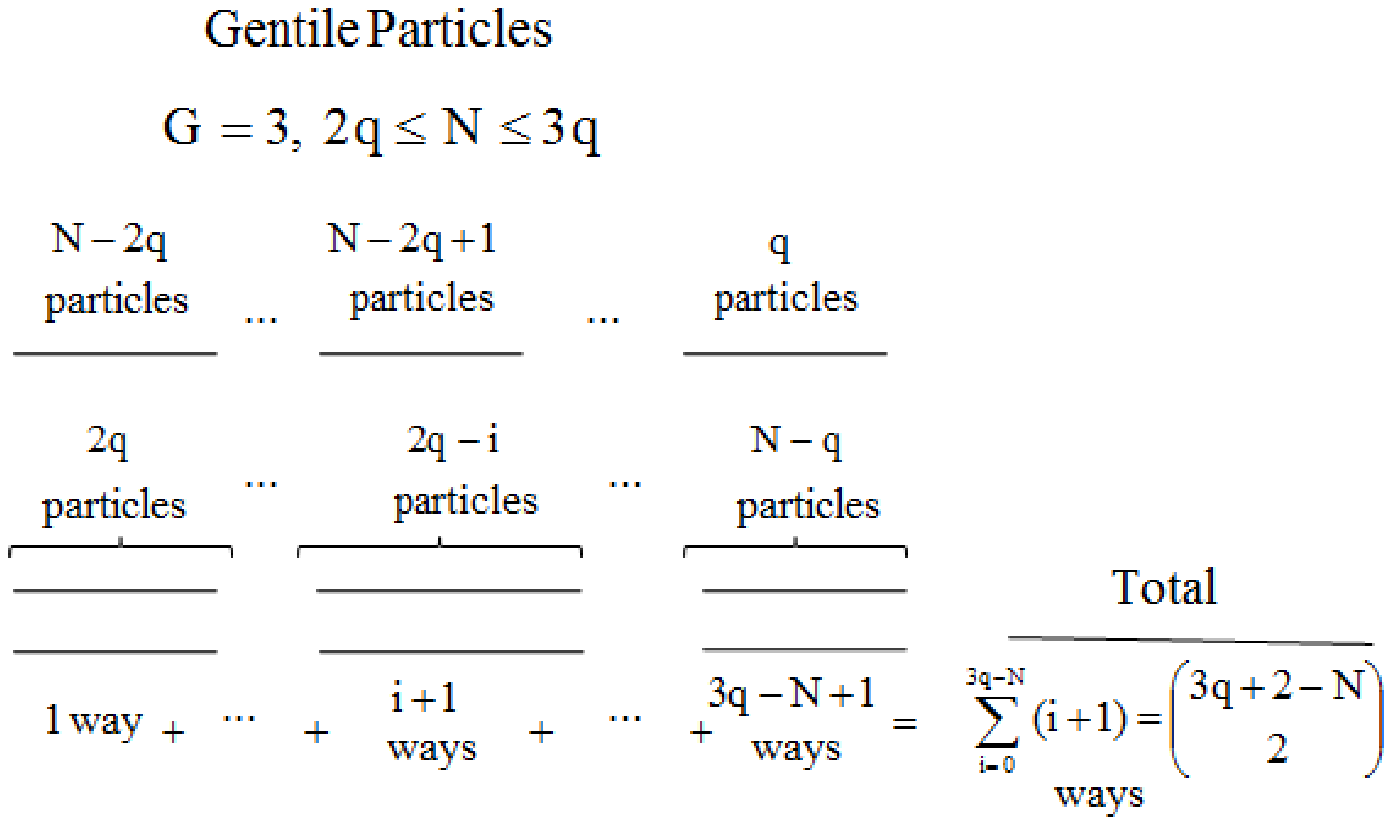}
\caption{Statistical weight of N Gentile particles for three levels
$R_q(N,3)$ when $2q \leq N \leq 3q $} \label{gonuca}
\end{figure}

We first begin with the case $2q \leq N \leq 3q$. We will use the
first equality in Eq. \eqref{goniki} for $R_q(2,N)$, because the
total number of particles in the levels 2 and 3 is more than $q$ for
this case, as mentioned above. One can put at least $(N-2q)$ and at
most $q$ particles to level $1$ . If there are $(N-2q)$ particles in
level 1, one finds from Eq. \eqref{goniki} that there are only
$2q+1-2q=1$ way to distribute remaining $2q$ particles into two
levels. If there are $i$ particles in level 1, there are
$[2q+1-(N-i)]$ ways of distributing remaining $(N-i)$ particles to
the remaining two levels (see Figure \ref{gonuca}). Hence for
$2q\leq N\leq 3q$ one obtains
\begin{equation}
R_q(3,N)=\sum_{i=N-2q}^{q}2q+1-(N-i)=\left(
\begin{array}{c} 3q+2-N
\\ 2
\end{array}
\right) \quad  2q\leq N\leq 3q .\label{gonuc}
\end{equation}
For $q<N<2q$ one can use the same logic but one must be careful.
Because now it is possible to put $0$ to $q$ particles to level 1
and the formula will change depending whether there are less than
$N-q$ particles in level $1$ or more than $N-q$ particles in level
$1$. Because if there less than $(N-q)$ particles in level $1$ there
are more than $q$ particles in the remaining two levels and one uses
the first expression in Eq. \eqref{goniki} for $R_q(2,N)$. If there
are more than $N-q$ particles in level $1$, there are less than $q$
particles in the remaining two levels and one uses the second
expression in \eqref{goniki} for $R_q(2,N)$. So, $R_{q}(3,N)$ for
$q<N<2q$ is
\begin{eqnarray}
R_q(3,N)=\sum_{i=0}^{N-q}[2q+1-(N-1)]+\sum_{i=N-q}^{q}(N-i+1)\\
\nonumber \frac{(q+1)(2N+2-3q)+2(2q-N)(N+1)}{2}. \label{wgoniki}
\end{eqnarray}
As one may notice this formula can not be written as a combinatoric
formula. Therefore, we continue cases where $N$ is between $(G-1)q$
and $Gq$, first. If the number of levels is $4$ and if there are
$(3q\leq)N(\leq 4q)$ particles to distribute to these $4$ levels
there can be at least $N-3q$, at most $q$ particles in level 1.
Using the  Eq. \eqref{gonuc} one can easily find the number of ways
for distributing the remaining particles to the remaining three
groups and summing these results one gets
\begin{eqnarray}
R_q(4,N)=\sum_{i=N-3q}^{q} \left( \begin{array}{c} 3q+1-(N-i)
\\ 2
\end{array} \right)= \left( \begin{array} {c} 4q+3-N
\\ 3
\end{array} \right) \;\;\;\;3q\leq N\leq 4q .
\label{wgondort}
\end{eqnarray}
From the formulae \eqref{goniki},\eqref{gonuc} and \eqref{wgondort}
we propose that for $(G-1)q \leq  N \leq Gq$,
\begin{equation}
R_q(G,N)=\left( \begin{array}{c} Gq+G-1-N
\\ G-1
\end{array}
\right). \label{gongenel}
\end{equation}
Assuming Eq. \eqref{gongenel} is true, it is easy to prove
$R_q(G+1,N)=\left( \begin{array}{c} (G+1)q+G-N
\\ G
\end{array} \right)$.
This is done by the logic we have used for all cases until now: We
separate level $1$ from other $G$ levels, then find for all number
of allowed number of particles $(i)$ in level $1$ the number of
different ways to distribute remaining $(N-i)$ particles to the
remaining $G$ levels $R_q(G,N-i)$ using Eq. \eqref{gongenel} and
finally sum $R_q(G,N-i)$ for all allowed $i$. Since we are, for now,
interested in the case $Gq\leq N\leq(G+1)q$, there can be at least
$N-Gq$, at most $q$ particles in level 1. Thus
\begin{eqnarray}
& &R_q(G+1,N)=\sum_{i=N-Gq}^{q} \left( \begin{array}{c}
Gq+G-1-(N-i)
\\ G-1
\end{array} \right)= \left( \begin{array}{c} (G+1)q+G-N
\\ G
\end{array} \right) \nonumber \\ & &\textrm{where~} Gq\leq N\leq (G+1)q
\label{wgonGp1}
\end{eqnarray}
This formula is the same as the Eq. \eqref{gongenel} for $G$
replaced by $G+1$. Thus, we have shown by induction that the Eq.
\eqref{gongenel} gives the statistical weight for Gentile particles
when $(G-1)q \leq N \leq Gq$.

One can see from the Eq.(9) if $N<(G-1)q$ for given $G$ and $q$,
it is not possible to find a combinatoric formula for the
statistical weight of Gentile particles. However, the inductive
method can still be used. In this case, one can derive an
recursive formula using the inductive method. In order to find
$R_q(G,N)$ for $N<(G-1)q$, we again separate level $1$ from the
remaining $G-1$ levels. Since $N$ is less than $(G-1)q$, at least
$0$ bosons, at most $q$ bosons may occupy level
$1$\footnote{Recall that if $N>(G-1)q$ the minimum number of
particles that can occupy a level is $N-(G-1)q$, because at most
$(G-1)q$ particles are allowed to occupy remaining $(G-1)$
levels.}. If there are $i$ particles in level $1$, there are $N-i$
particles the remaining $G-1$ levels and there are $R_q(G-1,N-i)$
ways to distribute these particles to the remaining $G-1$ levels.
Therefore we can write
\begin{equation}
R_q(G,N)=\sum_{i=0}^{q}R_q(G-1,N-i) . \label{geksibirneksii}
\end{equation}
We know $R_q(G,N)$ for small $N$ values: $R_q(G,N)$ is the same as
the statistical weight for bosons if $N \leq q$, that is
\begin{equation}
R_q(G,N)=\left( \begin{array}{c} N+G-1
\\ N
\end{array}
\right)\;\;\;\;\;\;\;\; N\leq q. \label{nkucukqoldugunda}
\end{equation}
So, for a given $q$, beginning from the statistical weights of small
$G$ and $N$ values and utilizing Eq. \eqref{nkucukqoldugunda} when
possible, it is easy to calculate $R_q(G,N)$ recursively, by means
of a computer program.  In the next section, we will first compare
this recursive formula with the recursive formula obtained by Perez
and Tun (the Eq. (4) in \cite{pereztun}).

The inductive method developed here is also applicable to the
statistics constructed in \cite{xie} by Xie and Dai which is more
general than Gentile statistics. In this statistics the value of
the variable q that is the maximum number of particles that can
occupy a state is not constant but may change in other words also
the value of the q is state dependent. Therefore we label the
maximum number of particles for different state by $q_i$ where
$i=1,2,..G$. The order of labelling is not important in the
calculation of the statistical weight. Therefore we arrange the
states such that the inequalities
\begin{equation}
q_1\leq q_2 \leq \ldots \leq q_G
\end{equation}
are satisfied. In this case the maximum number of particles that
can be distributed to the $G$ levels are
\begin{equation}
N_{max}=\sum_{i=1}^{G} q_i \label{Nmax}
\end{equation}
We will first show that the combinatoric formula given in Eq.
\eqref{gongenel} can be extended for this case if the total number
of particles $N$ satisfies the inequality $N_{max}-q_1 \leq N \leq
N_{max}$.

We will begin  with two states as in the case of Gentile
statistics. We assume that there are two states. These states can
be occupied by at most $q_1$ and $q_2$ particles and we order them
such that $q_1 \leq q_2$. If $N \leq q_1$ the distribution is
similar to the bosonic case and thus $R_{\{q\} }(G,N)=N+1$
\footnote{For the Xie-Dai distribution we denote the statistical
weight as $R_{\{q\} }(2,N)$ since the value of q is not constant.
We denote by $ \{q\}=q_1,q_2,\ldots,q_G $} . If $q_1 \leq N \leq
q_2$ then there can be at least $N-q_1$ and at most $N$ particles
in the second state. Therefore $R_{\{q\} }(2,N))=q_1+1$. If $q_2
\leq N \leq N_{max}=q_1+q_2$ then there can be at least $N-q_1$
and at most $q_2$ particles in the state two and thus $R_{\{q\}
}(2,N)=N_{max}-N+1$. We can summarize the results obtained for the
statistical weight $R_{\{q\} }(2,N)$ as
\begin{equation}
R_{\{q\} }(2,N)= \begin{cases} N_{max}+1-N=\left( \begin{array}{c}
N_{max}+1-N
\\ 1
\end{array} \right)& q_2 \leq
N \leq N_{max} \;\; (a)
\\
q_1+1 & q_1 \leq N \leq q_2 \;\;\;\; \quad (b)
\\ N+1&  N \leq q_1 \;\;\;\;\,\qquad  \quad (c)
\end{cases}
\label{xDG2}
\end{equation}

Using the results obtained for two states, we will find a
combinatoric formula for three states if $N_{max} -q_1 \leq N \leq
N_{max}=q_1+q_2+q_3 $. (For the other cases it is not possible to
find a combinatoric formula and we will derive a recursive formula
as in the Gentile statistics later.) Given condition $N_{max} -q_1
\leq N$ and the fact that at most $q_3$ particle can occupy the
state 3, there can be at least $N-q_3$ particles in the state 3.
Because we assume that the lower limit for the total number of
particles is $N_{max}-q_1=q_2+q_3$, $N-q_3$ is larger or equal to
$q_2$ and we can use the condition (a) in the Eq. \eqref{xDG2} for
calculating the total number of different ways of distributing the
remaining particles to the remaining two states. If there are $i$
particles in the state 3 there will be $N-i$ particles in the
remaining two states and hence we get after some elementary
calculations
\begin{eqnarray}
R_{\{q\} }(3,N) &=& \sum_{i=N-(N_{max}-q_3)}^{q_3} \left[
N_{max}-q_3-(N-i)+1 \right]   \nonumber \\ &=& \left(
\begin{array}{c} N_{max}+2-N
\\ 2
\end{array}
\right) \;\;\;\textrm{for~}N_{max}-q_1 \leq N\leq N_{max}.
\label{xDG3}
\end{eqnarray}

As in the case of Gentile statistics using the part (a) of Eq.
\eqref{xDG2} and Eq. \eqref{xDG3} we propose that the formula for
a general number of states $G$ is
\begin{equation}
R_{\{q\} }(G,N) = \left(
\begin{array}{c} N_{max}+G-1-N
\\ G-1
\end{array}
\right) \;\;\;\textrm{for~}N_{max}-q_1 \leq N\leq N_{max}.
\label{xDGG}
\end{equation}
where $N_{max}$ is given by Eq. \eqref{Nmax}. Then assuming the
Eq. \eqref{xDGG} is valid we will get the same formula $G+1$
states. In this case $N_{max}=\sum_i^{G+1} q_{i}$. We assume again
that the total number of particles satisfy $N_{max}-q_1 \leq N
\leq N_{max}$. There can be at most $q_{G+1}$ particles in the
state $G+1$. Since $ N \geq N_{max}-q_1=q_2+ \ldots +q_{G+1}$ the
number of particles in the remaining states always satisfy the
necessary inequality given for Eq. \eqref{xDGG}. Since there can
be at least $ N-N_{maxG}$ and at most $q_{G+1}$ particles in the
state $G+1$ and we assume Eq. \eqref{xDGG} is valid we get
\begin{eqnarray}
R_{\{q\} }(G+1,N) &=& \sum_{i=N-N_{maxG}}^{q_{G+1}}  \left(
\begin{array}{c} N_{maxG}+G-1-(N-i)
\\ G-1
\end{array}
\right) \nonumber \\ &\textrm{for~}& N_{max}-q_1 \leq N\leq
N_{max}.
\label{xDGGp1}
\end{eqnarray}
where $N_{maxG}=\sum_{i=1}^{G} q_i$. Changing the dummy index as
$k=i-(N-N_{maxG})$ and using $\sum_{j=m}^n\left(
\begin{array}{c} j
\\ m
\end{array} \right)=\left( \begin{array}{c} n+1
\\ m+1
\end{array} \right)$ for $n \geq m$ we get
\begin{eqnarray}
R_{\{q\} }(G+1,N) &=& \sum_{k=0}^{N_{max}-N}  \left(
\begin{array}{c} k+G-1
\\ G-1
\end{array}
\right)=\sum_{k=G+1}^{N_{max}-N+G-1}  \left(
\begin{array}{c} k
\\ G-1
\end{array}
\right) \nonumber \\ & = &\left(
\begin{array}{c} N_{max}-N+G
\\ G
\end{array}
\right) \;\, \textrm{for~} N_{max}-q_1 \leq N\leq N_{max}
\label{xDGGp12}
\end{eqnarray}
Since this equation is the same equation with Eq. \eqref{xDGG} for
$G$ is replaced by $G+1$, we conclude that the statistical weight
for Xie-Dai distribution is given by Eq. \eqref{xDGG} if
$N_{max}-q_1 \leq N \ leq N_{max}$. Note that if all $q_i$s are
equal to each other Then $N_{max}=Gq$ and Eq. \eqref{xDGG} reduces
Eq. \eqref{gongenel} as it must be.

Now we will derive a recursive formula for Xie-Dai statistics
because like in the case of the Gentile statistics it is not
possible to derive a combinatoric formula if $q_1 < N <
N_{max}-q_1$. The recursive formula we will derive is valid for
all possible cases.  We will again use induction and separate the
state G from the other $G-1$ states. If $ N > N_{max}-q_1 $ there
can be at least $N-N_{max}$ particles and if $N \leq N_{max}-q_1 $
there can be at least 0 particles in the state $G$. If $N>q_G $
then there can be at most $q_G$ but if $q_G<N$ there can be at
most $N$ particles in the state $G$. Therefore we define
$b=max[N,q_G]$ and $t=min[0,q_G]$ where $max[a,b]$ and $min[c,d]$
denote the larger one of the numbers $a$ or $b$; and the smaller
one of the numbers $c$ or $d$, respectively. As in the case of the
Gentile statistics, we realize that if there are $i$ particles in
level $G$, there are $N-i$ particles the remaining $G-1$ levels
and there are $R_{\{q\}}(G-1,N-i)$ ways to distribute these
particles to the remaining $G-1$ levels. Therefore we can write
\begin{equation}
R_{\{q\}}(G,N)=\sum_b^t R_{\{q\}}(G-1,N-i) \, .
\label{xDgen}
\end{equation}
This formula is valid for all possible cases. We will show the
change of $R_{\{q\}}(G,N)$ with $N$ and $G$ for different cases in
the following section.

%%%%%%%%%%%%%%%%%%%%%%%%%%%%%%%%%%%%%%%%%
\section{Results and Discussion}
In the reference \cite{pereztun}, the authors compared statistical
weight of  the Gentile distribution with the statistical weights
of the Wu and Polychranokos statistics. We will not repeat these
comparisons here.
\begin{figure}[t!]
\includegraphics[height=5cm, width=8cm]{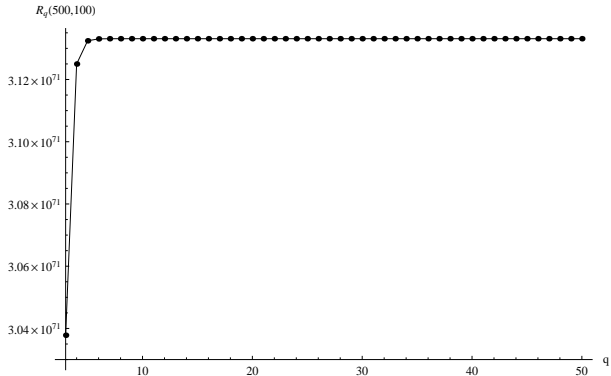}
\caption{The statistical weight for Gentile particles when $N=50$,
$G=500$ vs. the maximum occupation number q. q changes from $3$ to
$50$. The dots  shows the results of the recursive formula derived
in \cite{pereztun}, and the smooth curve show the results of the
recursive formula \eqref{geksibirneksii} obtained in the previous
section.} \label{distcomp}
\end{figure}
However, we will show in Figure \ref{distcomp} that the recursive
formulae derived by Perez and Tun, which is
\begin{figure}[b!]
\includegraphics[height=5cm, width=12cm]{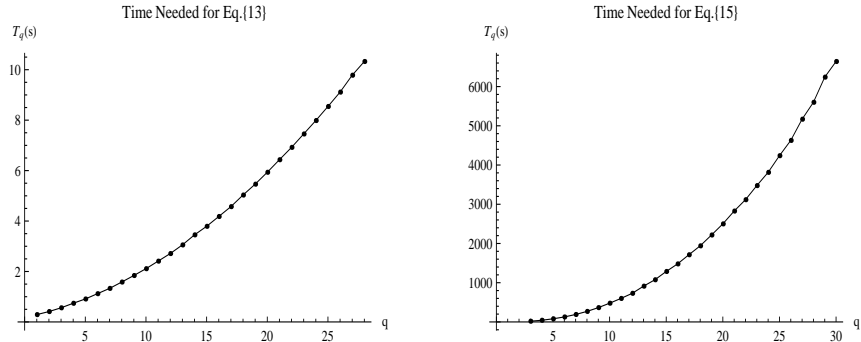}
\caption{The time in seconds needed by a mathematica program to
calculate $R_q(100 ,100 \,q) $ vs. the maximum occupation number
q. The figure on the left shows the time for the recursive formula
\eqref{geksibirneksii} obtained in the previous section and the
figure on the right shows the time for the recursive formula
derived in \cite{pereztun}. } \label{timecomp}
\end{figure}
\begin{equation}
R_q(G,N)=\sum_{j=0}^{[N/q]} \left( \begin{array}{c} G \\ j
\end{array}  \right) R_{q-1}(G-j,N-q j) \label{pereztun}
\end{equation}
and the recursive formula we have derived in Eq.
\eqref{geksibirneksii} is equivalent . In this figure, the change
of $R_q(500,50)$ with respect to $q$ is shown calculated by using
the recursive formula \eqref{geksibirneksii} and \eqref{pereztun}.
One can see from the Figure \ref{distcomp} that the result of
these formulae coincide with each other. One can also calculate
for specific values and see that both formulae give exactly the
same result.

Now we will compare the time  needed for finding the statistical
weight using the recursive formulae found in this study and in Ref.
\cite{pereztun}. We present in Figures \ref{timecomp}, the time in
seconds needed by a Mathematica program  for finding the statistical
weight $R_q(100, 100 \, q)$ (that is the statistical weight of $q$
times 100 Gentile particles distributed to $G=100$ levels) using the
recursive formula found by the inductive method presented in the
last section and using the recursive formula given in Ref.
\cite{pereztun}, respectively. One can see from these figures the
formula presented in this paper is approximately $600$ times faster
than the corresponding formula derived in \cite{pereztun} for a
given q.

One can see why the recursive formula we have derived, Eq.
\eqref{geksibirneksii} in the previous section produce the result of
the statistical weight faster for a given $q$  than the formula
\big(Eq. \eqref{pereztun} \big)) found in \cite{pereztun}. When one
uses the Eq. \eqref{pereztun} for finding $R_q(G,N)$ one needs the
statistical weights with smaller $q$ values. However, in the
recursive formula given in Eq. \eqref{geksibirneksii} one needs only
the statistical weights for smaller $N$ and $G$ values for a given
$q$. Therefore we may conclude that if the system has a determined q
like in N-annules the recursive formula \eqref{geksibirneksii} is
more useful than \eqref{pereztun}. However, if there are systems
which may have not a constant but a changing $q$ values than the
formula \eqref{pereztun} obtained in \cite{pereztun} is advantageous
compared the formula we have obtained.

\begin{figure}[t!]
\hskip -1cm
\includegraphics[height=4cm, width=14cm]{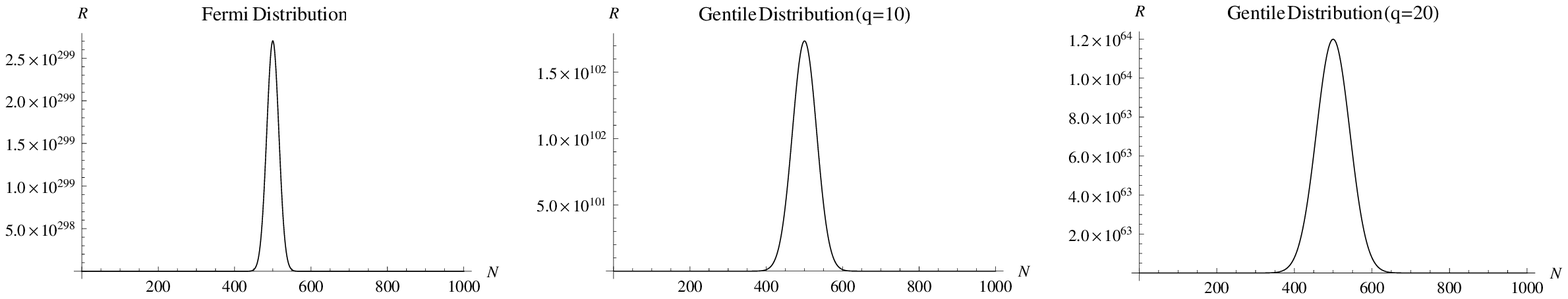}
\caption{The comparison of the statistical weights  for the Fermi
distribution for $G=1000$ and for the Gentile distributions for
$q=10$, $G=100$; $q=20$, $G=50$ with respect to number of
particles.} \label{comp3}
\end{figure}
Now, we compare the change of statistical weights for the Fermi
distribution for $G=1000$ and for the Gentile distributions for
$q=10$, $G=100$; $q=20$, $G=50$  with respect to the number of
particles. We choose the values $q$ and $G$ in Gentile
distributions and $G$ in Fermi distribution such that the maximum
number of particles ($N_{max}$) is $1000$ for all cases. As one
can see from the Figure \ref{comp3} the peak value occurs at
$N=500$, which is the half of $N_{max}$ for all cases. However the
peak is sharper in fermion case ($q=1$), and the peak is
broadening when $q$ increases. Thus, it is possible to conclude
that for Gentile particles with a large $q$ the steepest-descent
method, which are widely used for calculating the partition
function for distributions with sharp peaks(see e.g.
\cite{bkilic}), may not be used.

\begin{figure}[t!]
\hskip 1cm
\includegraphics[height=4cm, width=10cm]{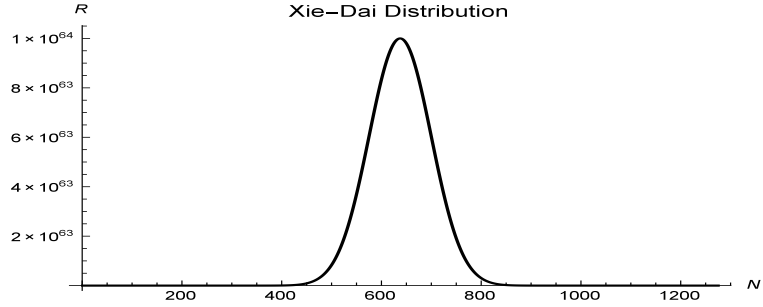}
\caption{The statistical weight for the Xie-Dai distribution with
respect to the total number of particles when there are $G=50$
states. We take $q_1=1, q_2=2, \ldots,q_{49}=49,q_{50}=50$.}
\label{XDF1}
\end{figure}

\begin{figure}[t!]
\hskip 1cm
\includegraphics[height=4cm, width=10cm]{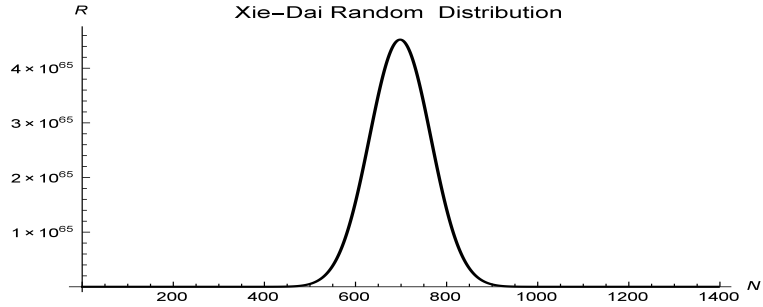}
\caption{The statistical weight for the Xie-Dai distribution with
respect to the total number of particles when $q_i$s are
determined randomly.} \label{XDF2}
\end{figure}

\begin{figure}[h!]
\hskip 1cm
\includegraphics[height=4cm, width=8cm]{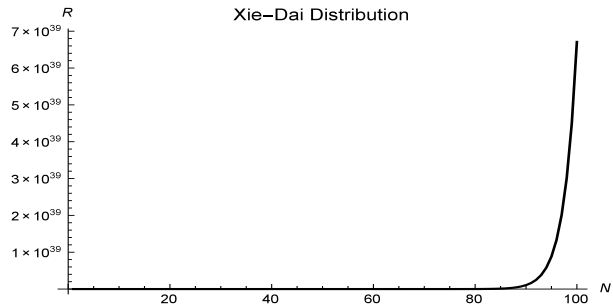}
\caption{The statistical weight for the Xie-Dai distribution with
respect to the total number of particles when the ground state is
bosonic and all the other states are fermionic.} \label{XDF3}
\end{figure}

Finally, we will show the change of the statistical weight for
Xie-Dai distribution with respect to the total number of particles
for different cases using the recursive formula in Eq.
\eqref{xDgen}. First we show the change of $R_{\{q\}}(50,N)$ with
$N$ where we take $G=50$ and $q_1=1, q_2=2,
\ldots,q_{49}=49,q_{50}=50$, that is the maximum number of
particles in the state 1 is 1 and it increases successively for
the following states. Since the total number of particles is
limited by $N_{max}=\sum_{i=1}^{50} i=1275$ for this case, we
obtain a symmetric distribution with respect to $N_{max}/2$ as
shown in Figure \ref{XDF1}.
Then, we have determined the maximum number for different states
randomly and calculated the statistical weight using Eq.
\eqref{xDgen} for different values. In this case the total number
of particles are again limited by $N_{max}=\sum_{i=1}^{50} q_i$.
We present the change of $R_{\{q\}}(50,N)$ with respect to $N$ for
this case in Figure \ref{XDF2}.
Finally, we investigate the case studied by Xie and Dai in
\cite{xie}. In this study the authors take the first state as
bosonic and the other states as fermionic. That is the number of
particles are not limited for the ground state but only one
particle can occupy the remaining excited states. In this case the
maximum number of particles is not limited and the statistical
weight is monotonically increasing with $N$. The change of
$R_{\{q\}}(50,N)$ with $N$ is shown in Figure \ref{XDF3}.
There are almost infinite number of different cases one can
investigate for Xie-Dai distribution and we have studied the
statistical weight only for three different cases. However, one
can use Eq. \eqref{xDgen} for all possible cases.
%%%%%%%%%%%%%%%%%%%%%%%%%%%%%%%%%%%


\begin{thebibliography}{99}
%
\bibitem{pereztun} R. Hernandez- Perez, D. Tun Physica A 384 (2007)
297.
%
\bibitem{xie} W. S. Dai, M. Xie, J. Stat. Mech. (2009) P07034.
%
\bibitem{avinashkhare} A. Khare, Fractional Statistics and Quantum Theory, World Scientific, Singapore, 2.ed  (2005)
%
\bibitem{leinaas} J.M. Leinaas, J. Myrheim, Nuovo Cimento 37B (1977).
%
\bibitem{wilczek} F. Wilczek (Ed.), Fractional Statistics and Anyon Superconductivity, World Scientific, Singapore, 1989.
%
\bibitem{nayak} C. Nayak, F. Wilczek, Phys. Rev. Lett. 73 (1994) 2740.
%
\bibitem{haldene} F.D.M. Haldene Phys. Rev. Lett. 67 (1991) 937.
%
\bibitem{wu} Y.S. Wu Phys. Rev.Lett. 73 (1994) 922.
%
\bibitem{polychranokos} A.P. Polychronakos  Phys. Lett. B 365 (1996) 202.
%
\bibitem{gentile} G. Gentile, Nuovo Cim., 17 (1940) 493.
%
\bibitem{dutt} R. Dutt, A. Gangopadhyaya, A. Khare, U.P. Sukhatme,
Int J. Mod. Physics A 9 (1994) 2687.
%
\bibitem{chaturvedi} S. Chaturvedi, V. Srinivasan, Physics A 246
(1997) 576.
%
\bibitem{bysto} A.G. Bysto, Nuclear physics B 604 (2001) 455.
%
\bibitem{daiannals} W. S. Dai, M. Xie, Annals of Physics 309 (2004)
295.
%
\bibitem{daiphysics} W. S. Dai, M. Xie, Physics Letter A 373 (2009)
1524.
%
\bibitem{macdonald} Z. MacDonald, B.P. van Zyl, Journal of physics:
Mathematical and Theoretical 46 (2013) 045001.
%
\bibitem{sriatson}  C.S. Srivatsan, M.V.N. Murthy, R.K. Bhaduri,
Pramana-Journal of physics 66 (2006) 485.
%
\bibitem{niven} R.K. Niven, Eur. Phys. J.B 70 (2009) 49.
%
\bibitem{mirza} B. Mirza, H. Mohommadzeh, Phys. Rev. E 82 (2010)
031137.
%
\bibitem{melijanac} S. Meliajanac, M. Milekovic, M Stojic, J. Phys
A.: Mathematical and General, 32 (1999) 115.
%
\bibitem{xie2} W. S. Dai, M. Xie, J. Stat. Mech. (2009)  P04021.
%
\bibitem{HP} T. Holstein, H. Primakoff, Phy. Rev. 58 (1940) 1098.
%
\bibitem{xie1} W. S. Dai, M. Xie, Annals of Physics 332 (2013)
166.
%
\bibitem{grosso} G. Grosso, G.P. Parraviccini, Solid State
Physics, Academic Press, San Diego, 2.ed (2003).
%
\bibitem{auccaise} R. Auccaise, J. Teles, T.J. Bonagamba, I.S.
Oliverira, E.R. deAzevedo, R.S. Sarthour, The Journal of Chemical
Physics 130 (2009) 144501.
%

\bibitem{yoshen} Yao Shen and B.Y. Jin, The Journal of Physical
Chemistry A 117 (2013) 12540.
%
\bibitem{bkilic} F. B\"{u}y\"{u}kk\ti l\ti \c{c}, H. Uncu, D.
Demirhan, Eur. Phys Jour. B. 35 (2003) 111.

\end{thebibliography}
\end{document}